\documentclass[11pt,a4paper]{article} 
\usepackage{jcappub}

\usepackage{amsmath}
\usepackage{amsfonts}
\usepackage{graphicx}
\usepackage{hyperref}
\usepackage{natbib}
\usepackage{bm}
\usepackage{tcolorbox}
\usepackage{amssymb}
\usepackage{mathbbol}
\usepackage{comment} 

\usepackage{cleveref}

\renewcommand{\d}{\mathrm{d}}
\newcommand{\e}[1]{\mathrm{e}^{{#1}}}
\newcommand{\im}{\mathrm{i}}
\newcommand{\vect}[1]{\bm{\mathrm{{#1}}}}

\newcommand{\curlD}{\mathcal{D}_t}
\newcommand{\curlDN}{\mathcal{D}_N}
\newcommand{\curlDplain}{\mathcal{D}}
\newcommand{\curlDtau}{\mathcal{D}_{\tau}}
\newcommand{\curlDtauprime}{\mathcal{D}_{\tau'}}

\newcommand{\Mp}{M_{\mathrm{P}}}
\newcommand{\GeV}{\text{GeV}}

\newcommand{\para}[1]{\par\vspace{2.5mm}\noindent\textbf{{#1}.---}}
\newcommand{\paranodot}[1]{\par\vspace{2.5mm}\noindent\textbf{{#1}---}}

\DeclareMathOperator{\PathOrder}{\mathsf{P}}

\title{Computing observables in curved multifield models of inflation \\
{\Large\textmd{\textsl{A guide (with code) to the transport method}}}}
\author[1]{Mafalda Dias,}
\author[2,3]{Jonathan Frazer,}
\author[1]{David Seery}

\affiliation[1]{\footnotesize Astronomy Centre, University of Sussex, Brighton BN1 9QH, United Kingdom}
\affiliation[2]{\footnotesize Department of Theoretical Physics, University of the Basque Country, UPV/EHU, 48040 Bilbao, Spain}
\affiliation[3]{\footnotesize IKERBASQUE, Basque Foundation for Science, 48011 Bilbao, Spain}

\abstract{We describe how to apply the transport method to
compute inflationary observables in a broad range of
multiple-field models.
The method is efficient and encompasses scenarios with
curved field-space metrics, violations of slow-roll conditions
and
turns of the trajectory in field space.
It can be used for an arbitrary mass spectrum,
including massive modes and models with quasi-single-field
dynamics.

\noindent
In this note we focus on practical issues.
It is accompanied by a \emph{Mathematica} code
which can be used to explore suitable models,
or as a basis for further development.}
 

\begin{document}

\maketitle
\flushbottom

\section{Introduction}
\label{sec:introduction}

Inflation \cite{Guth:1980zm,Linde:1981mu,Albrecht:1982wi} is a scenario for the very early universe
according to which
all large-scale structure
originated as quantum fluctuations.
This idea
has been tested by increasingly detailed observations
of anisotropies in the cosmic microwave background (CMB),
and
its broad predictions are
now known to be
compatible with
their measured statistical properties~\cite{1303.5082}.
This
is
a significant achievement
which symbolises the maturation of modern cosmology
into a precision science.
However,
despite this phenomenological success,
it remains unclear whether or not the microphysical origin of inflation can be understood. 

The simplest inflationary models comprise only
a single scalar degree of freedom,
taken to have a canonical kinetic term and 
a potential representing self-interactions.
Minimal models of this kind are sufficient to
obtain predictions consistent with observational constraints,
and therefore are natural from the perspective of simplicity
and economy.
From the perspective of fundamental physics, however, their status is uncertain.
If the inflationary scale is widely separated from the next
relevant mass scale
it may be natural to have only a single degree of freedom
which is light by comparison,
and a simple potential.
But if the inflationary scale is not far below the next relevant scale---%
in particular, if the ultraviolet completion of the theory
gives rise to multiple degrees of freedom which are light
compared to the inflationary scale---then
it may be that a model with \emph{several}
active scalar fields,
coupled through a nontrivial potential or field-space metric,
represents the most natural possibility.

Which of these choices is a better match for the
large-scale structure we observe in our universe
is a question to be resolved by measurement.
To do so will require
a clear understanding of the
qualitative predictions
which can be obtained in each scenario.
In single-field potential-dominated models,
an extended campaign of exploration
has provided guidance about what can be expected.
In this case the textbook approach to perturbations
is applicable.
Only a few e-folds around horizon exit are relevant;
the decoupling principle tells us that effects long before
horizon exit are negligible,
and long after it the perturbations become conserved.
During these few e-folds we can assume the Hubble rate to be nearly
constant and take 
the inflaton mass to be negligible.

In more complex models these simplifications no longer apply.
If the field-space trajectory exhibits turns
or other 
features 
around horizon exit
then
the calculation must usually begin
far inside the horizon,
tracking
effects from
perturbations
with masses around the Hubble scale.
Also, where multiple
fields remain relevant after horizon exit
we must continue to integrate
until their observable effects
are extinguished,
possibly much later in the inflationary era or
even long after it has ended.
All but the heaviest modes will be relevant after horizon exit,
including perturbations in the
momenta.

These complexities frustrate
the traditional textbook approach.
If only a few of them are present it may still
be possible to make analytic progress.
But more generally, in models where they all occur,
a numerical method is almost essential.

In this paper we show how
the complexities introduced by many relevant scales
can be accommodated, whether these scales
are associated with masses in the
particle spectrum
or curvature scales in the field-space manifold.
We illustrate our method with an explicit
\emph{Mathematica} implementation.%
    \footnote{This implementation was originally developed to study a model of
    D-brane inflation~\cite{1203.3792}.
    The precise model is described in Ref.~\cite{1103.2775}. The Lagrangian is
    $\mathcal{L}=a^{3}\big[\frac{1}{2}T_{3}G_{ij}\dot{\phi}^{i}\dot{\phi}^{j}-V(\phi)\big]$
    where $a$ is the scale factor and $T_{3}$ is the brane tension.
    It consists of six fields $\phi^i$ representing
    coordinates in the throat
    of a Klebanov--Witten geometry which
    can be described by a noncompact conifold built over
    the five-dimensional $\mathrm{SU}(2) \times \mathrm{SU}(2))/\mathrm{U}(1)$
    coset space $T^{1,1}$.
    The details of this
    geometry are encoded in the nontrivial field-space
    metric $G^{ij}$.
    The potential includes stochastic contributions from the bulk
    and it consists of $\sim 600$ terms.}
The version discussed here is available from \href{https://transportmethod.wordpress.com/}{transportmethod.com}.
We have attempted to simplify it as much as possible,
with the intention of making it easy to follow.
It can be used to compute the two-point function of
inflationary fluctuations in a model with any
number of fields and arbitrary potential and field-space metric.
The numerical method (to be described in \S\ref{sec:transport})
is efficient, and therefore despite being implemented in
\emph{Mathematica} it is fast enough for practical model exploration.
In situations where \emph{Mathematica}
is too slow or inconvenient,
such as Monte Carlo sampling,
it could serve as a reference implementation.%
	\footnote{Other general purpose codes exist.
	\href{http://theory.physics.unige.ch/~ringeval/fieldinf.html}{\emph{FieldInf}}
	is a Fortran code capable of computing
	the inflationary power spectrum in
	an $N$-field model with
	a nontrivial field-space metric~%
	\cite{Ringeval:2005yn,Martin:2006rs,Ringeval:2007am}.
	\href{http://modecode.org}{\emph{ModeCode}}
	and
	\href{http://modecode.org}{\emph{MultiModeCode}}
	are similar Fortran codes
	designed (respectively)
	for
	single- and multiple-field
	models.
	They are restricted to a trivial field-space
	metric but
	emphasize Monte Carlo sampling~%
	\cite{Mortonson:2010er,Easther:2011yq,Norena:2012rs,Price:2014xpa}.
	\href{http://pyflation.ianhuston.net}{\emph{Pyflation}}
	is a Python code which solves
	the Mukhanov--Sasaki equation to first-order
	in multiple-field models,
	and to second-order in single-field models~\cite{Huston:2009ac,Huston:2011vt,Huston:2011fr}.
	Like \emph{ModeCode} it is restricted to a trivial field-space metric.}

\para{Synopsis}%
This paper is divided into four principal parts.
In~{\S\ref{sec:transport}
we explain how to derive differential equations
which express the time evolution of
field-space correlation functions in the spatially flat
gauge.
We allow an arbitrary potential and
field-space metric.
With appropriate initial conditions this system of equations
can be used to capture contributions (including 
quantum effects) from all mass
scales as the fluctuations approach,
pass through, and eventually evolve outside the
Hubble length.
We discuss the selection of initial conditions
in~\S\ref{sec:IC}.

In~\S\ref{sec:gauge} we explain how
to relate the flat-gauge field-space correlation functions
to the statistical properties
of the density perturbation, which is the observable
quantity.

For a given model the major numerical
uncertainty is the duration of
nontrivial evolution on super-Hubble
scales.
In principle---%
no matter which
scheme we use to compute the properties
of observables---%
the equations for all inflationary perturbations
should be integrated
up to the last scattering surface, where they
supply initial conditions
for the cosmic microwave background anisotropies.
In practice this is very onerous,
and anyway would require us to integrate through epochs
of cosmological history, such as reheating,
about which we know nothing.
To evade both these issues we must usually rely on
the dynamics becoming `adiabatic' at some
point during inflation,
or not long after---meaning that
the isocurvature modes which can source time dependence of
the density perturbation become exhausted, and it ceases
to evolve.
In~\S\ref{sec:adiabatic} we discuss the issues which
arise when trying to detect whether 
this limit has been reached.

\para{Notation}%
We set $c = \hbar = 1$ and express the gravitational coupling
by the reduced Planck mass $\Mp^{-2} \equiv 8 \pi G$.
Greek indices from the beginning of the alphabet,
($\alpha$, $\beta$, \ldots)
label the species of light scalar fields;
Greek indices from the middle of the alphabet
($\mu$, $\nu$, \ldots)
label spacetime dimensions.
Spacetime indices are not needed except in Eq.~\eqref{eq:curved-action}.

\section{Transport equations for correlation functions}
\label{sec:transport}

An inflationary model with curved field space is governed by the action
\begin{equation}
    \label{eq:curved-action}
    S= \frac{1}{2} \int \d^3 x \, \d t \; \sqrt{-g}
    \Big\{ \Mp^2 R -
     G_{\alpha\beta}g^{\mu\nu}  \,  \partial_\mu \phi^\alpha \partial_\nu \phi^\beta-2V\Big\},
\end{equation}
where $G^{\alpha\beta}$ is the field-space metric,
$V$ is the potential
and $g^{\mu\nu}$ is the space-time metric.
At background level we take this to be Robertson--Walker,
\begin{equation}
    \d s^2 = - \d t^2 + a(t)^2 \d \vect{x}^2 ,
\end{equation}
where $a(t)$ is the scale factor.
Both $G_{\alpha\beta}$ and $V$ may be arbitrary functions of
the fields $\phi^\alpha$ provided they
are compatible with field configurations which realize
an inflationary epoch.
The equation of motion for the unperturbed background fields is
\begin{equation}
    \label{eq:KG}
    \curlD\dot{\phi}^\alpha + 3 H \dot{\phi}^\alpha + G^{\alpha\beta}V_{\beta} = 0,
\end{equation}
where $V_\beta \equiv \partial_\beta V$,
an overdot denotes partial
differentiation with respect
to cosmic time $t$ and $\curlD$ denotes a covariant
time derivative,
$\curlD X^\alpha \equiv \dot{X}^\alpha +
\Gamma_{\beta\gamma}^\alpha \dot{\phi}^\beta X^\gamma$.
The connexion $\Gamma^\alpha_{\beta\gamma}$ is the
Levi--Civita connexion compatible with $G_{\alpha\beta}$.

Our aim is to study quantum fluctuations in this model,
which
are characterized by correlation functions
of the independent degrees of freedom.
Their precise identity is influenced by
our choice of spacetime coordinates.
In this paper we define time $t$ so that slices of constant
$t$ have zero Ricci curvature, up to possible tensor modes
which we neglect.
In these coordinates
the independent degrees of freedom are
fluctuations $\delta\phi^\alpha$ in the fields,
and the correlation functions characterizing
quantum fluctuations
are the $n$-point functions
$\langle \delta\phi^\alpha(\vect{k}_1, t_1) \delta\phi^\beta(\vect{k}_2, t_2) \rangle$,
$\langle \delta\phi^\alpha(\vect{k}_1, t_1) \delta\phi^\beta(\vect{k}_2, t_2) \delta\phi^\gamma(\vect{k}_3, t_3) \rangle$,
and so on, together with their derivatives.
For applications to inflation we typically require only the equal-time
case where all $t_i$ are evaluated at some common point $t$.
The expectation value $\langle \cdots \rangle$ is taken in a state
which coincides with the Minkowski vacuum on deeply subhorizon scales.

\subsection{Capturing physical effects from all mass scales}
\label{sec:mass-capture}
In order to ensure that we
capture relevant physics from all mass scales,
we begin the calculation
sufficiently far inside the horizon
that vacuum initial conditions apply.
As we will show below,
in de Sitter space all degrees of freedom of fixed mass
become effectively \emph{massless} on subhorizon
scales,
so---irrespective of the mass spectrum---we
can obtain initial conditions for each $n$-point function
to arbitrary accuracy
by
beginning the calculation
sufficiently long before horizon exit. The details
are discussed in~\S\ref{sec:IC}.

We then apply the in--in formalism
to derive an evolution equation for each $n$-point
function, incorporating all masses
and (in principle) quantum effects.%
  \footnote{In principle the evolution equations contain
  the same information as the loop expansion of
  conventional perturbation theory,
  but not nonperturbative information
  such as instanton effects.
  In practice one must truncate each evolution equation,
  which is equivalent to truncating the
  loop expansion at a particular order.
  In this paper we work to tree level, which is
  already sufficient to capture those quantum interference effects
  around horizon crossing
  which determine the `quantum' part of the
  Feynman calculation~\cite{Lyth:2005fi,Seery:2005gb}.} 
These evolution equations
are equivalent to the separation into
in--out expectation values and subsequent
Feynman expansion
used by Maldacena and later authors
to obtain analytic estimates of
the correlation functions~\cite{Maldacena:2002vr}.
But unlike the expansion into diagrams they
do not involve Green's functions---only ordinary
differential equations.
Therefore they constitute a differential formulation
of the theory rather than an integral one,
and can be handled by conventional ODE solvers.

\para{Perturbed action}%
To second order in amplitude, the action governing
small fluctuations $\delta\phi^\alpha$
around a homogeneous solution
$\phi^\alpha(t)$
of~\eqref{eq:curved-action}
can be written~\cite{1208.6011,Sasaki:1995aw}
\begin{equation}
    \label{pertaction}
    S \supseteq \frac{1}{2} \int \frac{\d^3 k}{(2\pi)^3} \, \d t \; a^3
    \bigg\{
        G_{\alpha\beta}
        \big[ \curlD \delta\phi^\alpha(\vect{k}) \big]
        \big[ \curlD \delta\phi^\beta(-\vect{k}) \big]
        - \bigg(
            \frac{k^2}{a^2}G_{\alpha\beta}+ M_{\alpha\beta}
        \bigg)
        \delta\phi^\alpha(\vect{k}) \delta\phi^\beta(-\vect{k})
    \bigg\}
    ,
\end{equation}
where the effective mass matrix $M_{\alpha\beta}$ is defined by
\begin{equation} 
    M_{\alpha\beta}
    \equiv 
        V_{\alpha;\beta}
        - R_{\alpha\lambda\mu\beta} \dot{\phi}^\lambda \dot{\phi}^\mu
        - \frac{1}{a^3 \Mp^2} \curlD
        \bigg(
            a^3 \frac{\dot{\phi}_\alpha \dot{\phi}_\beta}{H}
        \bigg) .
\end{equation}
In this expression
$V_{\alpha;\beta} \equiv \partial_\beta V_{\alpha}
-\Gamma_{\alpha\beta}^{\gamma}V_{\gamma}$
is the covariant derivative of $V_{\alpha}$,
and $R_{\alpha\lambda\mu\beta}$ is the Riemann tensor built from the
metric connexion $\Gamma^\alpha_{\beta\gamma}$.
As before,
an overdot denotes a partial derivative with respect
to $t$.
Both $M_{\alpha\beta}$ and $G_{\alpha\beta}$ should be
evaluated on the homogeneous background
$\phi^\alpha(t)$.

Our formalism requires only
Eq.~\eqref{pertaction}.
It is not necessary that it derives from
an action of the form~\eqref{eq:curved-action}
which controls both the background and fluctuations.
It particular, it
may happen that~\eqref{pertaction} applies to the
fluctuations
in scenarios which have a more general noncanonical
kinetic structure
than~\eqref{eq:curved-action}.
(Note, however, that~\eqref{pertaction}
is not sufficiently general to cover fluctuations
in a $P(X,\phi)$ model
where the Lorentz symmetry between
time- and space-derivative terms would be broken by a
nontrivial sound speed $c_{\mathrm{s}}^2$.)
Where Eq.~\eqref{pertaction} applies, our evolution equations
for the two-point correlation functions apply likewise.
They may be used to compute the properties of the fluctuations,
although the background equations
[Eqs.~\eqref{eq:backg-dphi}--\eqref{eq:backg-dpi} below]
would require modification.

The constituents of
Eq.~\eqref{pertaction}, including the
perturbation $\delta\phi^\alpha$, transform covariantly
under a change of coordinates in field space.
This implies that $\delta\phi^\alpha$ must be understood
as a tangent vector, not a coordinate displacement.
The necessary formalism
underlying this interpretation
was given by Gong \& Tanaka~\cite{Gong:2011uw};
see also Ref.~\cite{1208.6011}.
To lowest order in $\delta\phi^\alpha$
this makes no difference, but it would become important
in any attempt to extend~\eqref{pertaction} to third order
or above.

\para{Quantization}%
To quantize the fluctuations we define
a momentum $\delta p_\alpha$ by the rule
$\delta p_\alpha = \delta S / \delta (\curlD \delta \phi^\alpha)$.
Then $\delta \phi^\alpha$ and $\delta p_\alpha$
are to be treated as operators satisfying the commutation relation
\begin{equation}
    [ \delta \phi^\alpha(\vect{k}_1) , \delta p_\beta(\vect{k}_2) ]
    = \im (2\pi)^3 \delta^\alpha_\beta \delta(\vect{k}_1 + \vect{k}_2)
\end{equation}
and the Hamiltonian is
\begin{equation}
    \label{eq:hamiltonian}
    \mathcal{H} = \int \frac{\d^3 k}{(2\pi)^3} \,
    \Big(
        \big[ \curlD \delta \phi^\alpha(\vect{k}) \big] \delta p_{\alpha}(-\vect{k})
        - \mathcal{L} 
    \Big),
\end{equation}
where $\mathcal{L}$ is the Lagrangian density appearing in~\eqref{pertaction}.
The equations of motion for $\delta\phi^\alpha$ and $\delta p_\alpha$ can be
determined from $\mathcal{H}$ via the Heisenberg equation.

In practice it is numerically more convenient to integrate in terms of
the e-folding number $\d N = H \, \d t$ rather than $\d t$ itself.
To do so at the level of the background
we define
\begin{equation}
    \label{eq:backg-dphi}
    \pi^\alpha \equiv \frac{\d \phi^\alpha}{\d N} = \curlDN \phi^\alpha ,
\end{equation}
in which it should be remembered that $\phi^\alpha$ (being a coordinate)
behaves like a field-space \emph{scalar},
whereas $\pi^\alpha$ (being the derivative of a coordinate)
behaves like a field-space \emph{vector}.
The background equations of motion now comprise~\eqref{eq:backg-dphi}
together with an evolution equation for $\pi^\alpha$,
\begin{equation}
    \label{eq:backg-dpi}
    \curlDN \pi^\alpha = (\epsilon-3) \pi^\alpha - \frac{G^{\alpha\beta} V_\beta}{H^2} .
\end{equation}
To effect a similar change for the quantized perturbations we
define
\begin{equation}
    \label{eq:delta-pi-def}
    \delta \pi_\alpha \equiv \frac{\delta p_\alpha}{H a^3}
    = \curlDN \delta \phi_\alpha ,
\end{equation}
where the index on $\delta\phi^\alpha$ should be lowered using the
metric $G_{\alpha\beta}$.
Because the operations of taking covariant perturbations
and covariant time-derivatives commute,
it follows that $\delta (\curlDN \phi^\alpha) = \curlDN \delta \phi^\alpha$,
and therefore $\delta \pi_\alpha$ can be regarded as an honest
perturbation to the rescaled background field~\eqref{eq:backg-dphi}~\cite{Gong:2011uw}.
This identification is not spoiled by raising or lowering the index because
the metric is covariantly constant.
The equations of motion for $\delta\phi^\alpha$ and $\delta\pi^\alpha$
can now be written
\begin{subequations}
\label{eq:dN-delta-group}
\begin{align}
    \curlDN \delta \phi^\alpha & =
        - \frac{\im}{H} [ \delta\phi^\alpha, \mathcal{H} ]
    \label{eq:dN-delta-phi}
    \\
    \curlDN \delta \pi^\alpha & =
        - \frac{\im}{H} [ \delta\pi^\alpha, \mathcal{H} ]
        + (\epsilon-3) \delta\pi^\alpha .
    \label{eq:dN-delta-pi}
\end{align}
\end{subequations}
Similar operator equations will hold in the quantum theory,
possibly modified by renormalizations
required to define composite operators
appearing in the commutators $[ \cdot , \mathcal{H} ]$.
Because these are operator equations they hold
for any insertion
of $\delta\phi^\alpha$ or $\delta\pi^\alpha$
in a correlation function,
provided it is not coincident with any other operator.
If we work only to tree-level the complexities
associated with renormalization
are not needed,
and we can work directly with the bare equations.
The noncanonical term in the evolution equation for
$\delta\pi^\alpha$
arises from the explicit time-dependent factors
$a^3$ and $H$ which appear in~\eqref{eq:delta-pi-def}.

\para{Transport equations}%
Salopek, Bond \& Bardeen
pointed out that
a single
solution of
the $2N$ differential equations~\eqref{eq:dN-delta-group}
is not sufficient to
compute the two-point correlation functions~\cite{Salopek:1988qh}.
A single solution
characterizes only how the late-time
$\delta\phi^\alpha$ and $\delta\pi^\alpha$
perturbations respond to a particular linear combination
of fluctuations at an earlier time,
and
to compute a correlation function we must know how the
late-time perturbations respond to an arbitrary early-time
perturbation.
This entails calculating $2N$ solutions of~\eqref{eq:dN-delta-group},
one for each independent initial condition.

Various formalisms exist to compute the required
solutions.%
	\footnote{Salopek, Bond \& Bardeen
	decomposed the late-time fluctuations into
	a linear combination of creation--annihilation
	operators for the early-time fields,
	and solved for the resulting mixing matrix~\cite{Salopek:1988qh}.
	See Ringeval~\cite{Ringeval:2007am},
	Huston \& Christopherson~\cite{Huston:2011fr} and
	Price, Frazer, Xu, Peiris \& Easther~\cite{Price:2014xpa} for recent applications.
	McAllister, Renaux--Petel and Xu solved Eqs.~\eqref{eq:dN-delta-group}
	explicitly, once for each independent initial condition~\cite{1207.0317}.
	Lalak, Langlois, Pokorski \& Turzynski applied a method
	very similar to that proposed by Salopek et al.~\cite{Lalak:2007vi}.
	The $\Gamma$-matrix or `propagator' introduced in Ref.~\cite{Seery:2012vj}
	is of a similar kind.
	Rigopoulos, Shellard \& van Tent~\cite{Rigopoulos:2004gr,Rigopoulos:2005xx}
	elaborated a formalism
	due to Groot Nibbelink \& van Tent~\cite{GrootNibbelink:2001qt},
	using a basis aligned with the
	instantaneous background trajectory. This can reduce the
	number of integrations required if we are prepared to give up knowledge
	of the isocurvature modes.
	A similar approach was used by Peterson \& Tegmark~\cite{Peterson:2010np,Peterson:2011yt}.}
We choose to use the operator
equations~\eqref{eq:dN-delta-group}
to obtain evolution equations for
each $n$-point function.
There is no loss of information compared with
solving for the field modes themselves, because
only the correlation functions
are meaningful:
the predictions of an inflationary model are
statistical, and are
obtained by interpreting the correlation
functions as ensemble averages and supposing
that our particular universe is typical.

In this paper we deal only with the
equal-time two-point functions,
which are sufficient to obtain lowest-order
inflationary observables.
There are four such functions:
$\langle \delta \phi^\alpha \delta \phi^\beta \rangle$,
$\langle \delta \pi^\alpha \delta \phi^\beta \rangle$,
$\langle \delta \phi^\alpha \delta \pi^\beta \rangle$,
and
$\langle \delta \pi^\alpha \delta \pi^\beta \rangle$.
To compress notation we denote a generic perturbation
such as
$\delta \phi^\alpha$
or
$\delta \pi^\alpha$
as $X^a$. The index $a$ ranges over the field and momentum
perturbations for each species $\alpha$.
To distinguish these
we continue to label field perturbations by
$\alpha$, $\beta$, \ldots,
but add a bar
to the species label for a momentum perturbation, giving
$\bar{\alpha}$, $\bar{\beta}$, \ldots, and so on.
A generic equal-time two-point function can now be written
\begin{equation}
    \label{eq:generic-2pf}
    \langle X^a(\vect{k}_1) X^b(\vect{k}_2) \rangle
    =
    (2\pi)^3 \delta(\vect{k}_1 + \vect{k}_2) \frac{\Sigma^{ab}}{k^3} .
\end{equation}
The evolution equation for $\Sigma^{ab}$ follows from
Ehrenfest's theorem,
\begin{equation}
    \label{eq:ehrenfest-theorem}
    \curlDN \langle X^a X^b \rangle
    =
    \langle ( \curlDN X^a) X^b \rangle
    + \langle X^a ( \curlDN X^b ) \rangle .
\end{equation}
This argument, valid for quantum-mechanical correlation functions,
was given by Mulryne~\cite{1302.3842}.
It entails solving of order $2N \times 2N$ differential equations
with a single initial condition, consistent with the
counting argument given above.
Using the symmetries of $\langle X^a X^b \rangle$ shows that there
are $3N(N+1)/2$ independent equations,
assuming that we use
the Weyl-ordered correlation functions to be defined in~\S\ref{sec:subhorizon}.

Up to this point our treatment has been exact, except that
by implicitly computing expectation values in a single state,
corresponding to a fixed field configuration,
we have restricted attention to what is visible in perturbation
theory in the vicinity of that configuration.
Therefore
Eqs.~\eqref{eq:dN-delta-group}
and~\eqref{eq:ehrenfest-theorem} contain the same information as
the loop expansion,
and
Eqs.~\eqref{eq:dN-delta-group}
must contain information about all orders in interactions,
represented by terms of all orders in $\delta\phi^\alpha$ and $\delta\pi^\alpha$
on the right-hand side.
Hence
\begin{equation}
    \label{eq:expansion-tensor}
    \curlDN X^a = {u^a}_b X^b + \cdots ,
\end{equation}
where ${u^a}_b$ is a matrix which can be computed%
    \footnote{As explained above,
    we are neglecting renormalization and operator mixing effects
    which may be generated by the proper definition of composite
    operators beyond tree level.}
using Eqs.~\eqref{eq:dN-delta-group}, and
`$\cdots$' denotes terms of order $X^a X^b$ and higher which we have
omitted.
Neglecting these terms corresponds to
working at tree-level in the loop expansion.
On the left-hand side, the covariant derivative $\curlDN$ acts on the
generic label $a$ appropriately for the `barred' and `unbarred'
types, so that
$\curlDN X^\alpha = \partial_N X^\alpha + \Gamma^\alpha_{\beta\gamma} \pi^\beta X^\gamma$
and
$\curlDN X^{\bar{\alpha}} = \partial_N X^{\bar{\alpha}} + \Gamma^\alpha_{\beta\gamma} \pi^\beta X^{\bar{\gamma}}$.

Combining Eqs.~\eqref{eq:generic-2pf}, \eqref{eq:ehrenfest-theorem}
and~\eqref{eq:expansion-tensor} gives
\begin{equation}
    \curlDN \Sigma^{ab} = {u^a}_c \Sigma^{cb} + {u^b}_c \Sigma^{ac} + \cdots ,
    \label{eq:transport-equation}
\end{equation}
which we describe as the \emph{transport equation}
for $\Sigma^{ab}$ \cite{0909.2256, 1008.3159}.
The curved field-space version derived here was first given in Ref.~\cite{1208.6011}.
Comparison with Eqs.~\eqref{pertaction}, \eqref{eq:hamiltonian}
and~\eqref{eq:dN-delta-group}
shows that the components of ${u^a}_b$ satisfy
\begin{equation}
\begin{split}
    {u^\alpha}_\beta & = 0 \\
    {u^\alpha}_{\bar{\beta}} & = \delta^\alpha_\beta \\
    {u^{\bar{\alpha}}}_\beta & = - \delta^\alpha_\beta \frac{k^2}{a^2H^2} - \frac{{M^\alpha}_\beta}{H^2} \\
    {u^{\bar{\alpha}}}_{\bar{\beta}} & = (\epsilon-3)\delta^\alpha_\beta .
\end{split}
\label{eq:u-tensor}
\end{equation}

\para{Mass dependence}%
The epoch of `horizon exit'
occurs when the physical
wavelength of order $a/k$ associated
with the comoving wavenumber $\vect{k}$
becomes comparable to the Hubble length $1/H$.
At this time the ratio $k/aH$ is of order unity.
Prior to horizon exit $k/aH \gtrsim 1$,
and provided we are not too close to the
start of the inflationary era
it will be possible to find
a point where $(k/aH)^2$ is much larger
than any component of the effective mass matrix ${M^\alpha}_\beta / H^2$.
If we choose to begin the calculation at or before this time
then all fields can be treated as effectively massless.
Where horizon exit is too close to the start of
inflation
it will not be possible to make ${M^\alpha}_\beta / H^2$
entirely negligible, and
we must find some other way
to supply initial conditions, presumably depending
on the pre-inflationary
history.
The calculation then becomes model dependent, but not harder
as a matter of principle. In this paper we will not
consider such possibilities.

If $H$ and ${M^\alpha}_\beta$ are nearly constant and
the components of ${M^\alpha}_\beta$ are
at most a few orders of magnitude larger than $H^2$
then the point where all fields become
effectively massless
might lie no more than
$N \gtrsim 3$ e-folds before horizon
exit.
At the other extreme, if
$H \approx 10^{12} \, \GeV$ (corresponding
to roughly GUT-scale inflation)
but ${M^\alpha}_\beta$
contains terms of order $\Mp$
then it could be necessary to begin
$N \gtrsim 14$ e-folds before horizon exit.
These estimates require refinement if $H$ or
${M^\alpha}_\beta$ vary significantly (see \S\ref{sec:howdeep} for a numerical prescription).
After horizon exit, $k/aH$ becomes exponentially small and
all but the most
suppressed contributions to ${M^\alpha}_\beta$ will be relevant.%
	\footnote{In certain models
	there may be
	a superheavy scale $\gg H$ above 
	which all modes can be neglected:
	the fluctuations in these modes
	decay rapidly because of their large mass.
	Also, the potential for a superheavy field
	is so steep that the background trajectory
	can be assumed to make no excursion
	in its direction.
	
	Although such a superheavy scale is normally
	assumed to exist,
	it has recently been appreciated that it is not
	straightforward to decide how large a mass is required
	before a field-space direction is negligible in this sense.
	The effect of massive modes is suppressed by inverse
	powers of the heavy mass $M$, but the rate of turn of the trajectory
	can be large. This gives a large number which can compensate
	for the smallness of $1/M$, making the direction more
	relevant than it would appear.
	A literature has developed to study these
	effects; for example, see
	Refs.~\cite{Achucarro:2012sm,Achucarro:2012yr}.}

Eqs.~\eqref{eq:transport-equation} and~\eqref{eq:u-tensor}
provide a unified way to study both
sub- and super-horizon regimes
while retaining all relevant contributions to ${M^\alpha}_\beta$.
In the literature these regimes are sometimes
associated with `quantum' and `classical' behaviour, but
both of these descriptions are marginally misleading.
In the subhorizon era we work only to tree level
and
therefore true quantum effects are absent,
but the initial conditions are quantum-mechanical
and mix growing- and decaying-mode
solutions for the elementary wavefunctions which
contribute to $\Sigma^{ab}$.
It is the interference between these modes which
determines the higher-order correlations
imprinted around the time of horizon exit.
In the superhorizon era
the evolution becomes classical in the restricted
sense that
decaying solutions die away.

\para{`In--in' and `$\delta N$' limits}%
Once suitable initial conditions have been selected,
it does not matter what spectrum of mass scales
exists in ${M^\alpha}_\beta$, or whether
$H$ varies significantly during or after the epoch
of horizon exit.
Eqs.~\eqref{eq:transport-equation}--\eqref{eq:u-tensor}
provide
an alternative
to the full diagrammatic description of the in--in formalism,
but one which is equivalent.
No further approximations are required.
To determine the evolution of each
correlation function we need only integrate
the transport equation.

When written in this form it is simple
to obtain the connexion between the in--in
formalism and the `separate universe picture',
which gives an intuitive classical
description of
the evolving fluctuations on superhorizon
scales~\cite{Starobinsky:1986fxa,Sasaki:1995aw,Wands:2000dp,
Rigopoulos:2003ak,Lyth:2004gb,Lyth:2005fi}.
In this limit the transport equation becomes
a Jacobi equation describing the dispersion of
neighbouring inflationary trajectories in field
space and can be integrated analytically to produce the
well-known `$\delta N$' Taylor expansion~\cite{Seery:2012vj, 1302.3842, 1208.6011}.%
    \footnote{It is possible, but substantially more complex, to see
    how the `$\delta N$' description emerges
    from the diagrammatic expansion~\cite{Dias:2012qy}.}

\para{Scale dependence of two-point function}%
A similar transport equation can be obtained for the
scale dependence of the 2-point correlation function,
which we measure using the matrix
\begin{equation}
    n^{ab} \equiv \frac{{\text d}\Sigma^{a b}}{{\text d} \ln k} .
\end{equation}
A transport equation for
$n^{ab}$ can be obtained by differentiating Eq.~\eqref{eq:transport-equation} \cite{1111.6544}
\begin{equation}\label{nab}
\curlDN n^{a b}= \frac{\d}{\d \ln k}\curlDN \Sigma^{a b}=u^a_{\	\, c}\	n^{c b} +u^b_{\	\, c}\	 n^{a c} + \frac{\d u^a_{\	\, c}}{\d \ln k}\Sigma^{c b}+ \frac{\d u^b_{\	\, c}}{\d \ln k}\Sigma^{a c}.
\end{equation}

\vspace{2mm}
\noindent\textbf{Tensor modes}%
\footnote{The calculations reported in this section were performed
in collaboration with Sean Butchers.}%
\textbf{.}---%
Tensor perturbations $\gamma_{ij}$ are transverse, traveless
perturbations of the spatial metric representing
gravitational waves.
Up to second order in amplitude their fluctuations are controlled
by the action
\begin{equation}
  S \supseteq \frac{\Mp^2 }{8} \int \d^3 x\, \d t \; a^3
    \bigg\{\dot{\gamma}_{ij}\dot{\gamma}_{ij}
               - 
            \frac{k^2}{a^2}\gamma_{ij}\gamma_{ij}
    \bigg\}
\end{equation}
where the Latin indices $i$, $j$ run over
the three spatial coordinates.
To obtain scalar equations it is convenient to decompose $\gamma_{ij}$
into a basis of polarizations.
In Fourier space this gives
\begin{equation}
  \gamma_{ij}(\vect{x}) =
    \int  \frac{\d^3 k}{(2\pi)^3} \, \sum_{s} \gamma_s(\vect{k}) e^s_{ij}(\vect{k}) \, \e{i\vect{k} \cdot \vect{x}}
\end{equation}
where the polarization sum $s$ runs over the orthogonal
states $s \in \{ +, \times \}$.
The corresponding polarization matrices are
traceless and satisfy $k_i e^s_{ij}(\vect{k}) = 0$,
and
are normalized so that
$e^s_{ij}e^{s'}_{ij}=2\delta^{ss'}$.
Each polarization $\Mp \gamma_s(\vect{k}) / \sqrt{2}$
behaves as a
canonically-normalized
free scalar field.
Therefore the two-point correlation function of tensor perturbations
is insensitive to the mass hierarchies of the system.

To write an evolution equation for it we define
the tensor momentum $\pi_s = \d \gamma_s/\d N$
and collect $\gamma_s$ and $\pi_s$ into
a two-component vector $\gamma_{s}^{\mathbb{a}} = ( \gamma_s, \pi_s )$.
The labels $\mathbb{a}$, $\mathbb{b}$, \ldots,
range over the tensor polarization and its momentum.
We write the two-point function of $\gamma_s^{\mathbb{a}}$ as
\begin{equation}
  \langle \gamma_{s}^{\mathbb{a}}(\vect{k}_1) \gamma_{s'}^{\mathbb{b}}(\vect{k}_2) \rangle
  = (2\pi)^3 \delta_{ss'} \delta(\vect{k} + \vect{k}') \Gamma^{\mathbb{a}\mathbb{b}} .
\end{equation}
It follows directly from Eqs.~\eqref{eq:transport-equation}
and \eqref{eq:u-tensor} that
$\Gamma^{\mathbb{a}\mathbb{b}}$ obeys the transport equation
\begin{equation}
  \frac{\d \Gamma^{\mathbb{a}\mathbb{b}}}{\d N} =
    {w^\mathbb{a}}_{\mathbb{c}} \Gamma^{\mathbb{c}\mathbb{b}}
    + {w^\mathbb{b}}_{\mathbb{c}} \Gamma^{\mathbb{a}\mathbb{c}} + \cdots  
\end{equation}
where (with no summation implied)
${w^{\gamma}}_{\gamma} = 0$,
${w^{\gamma}}_{\pi} = 1$,
${w^{\pi}}_\gamma = - k^2 / (a^2H^2)$
and
${w^{\pi}}_{\pi} = \epsilon-3$.

Following the same procedure that lead to Eq.~\eqref{nab}, it is straightforward to compute the scale dependence of the tensor spectrum. 
Defining the quantity
\begin{equation}
    {n_{\rm T}}^{\mathbb{a}\mathbb{b}} \equiv \frac{{\text d}\Gamma^{\mathbb{a}\mathbb{b}}}{{\text d} \ln k} ,
\end{equation}
it follows that its equation of motion is
\begin{equation}
  \frac{\d{n_{\rm T}}^{\mathbb{a}\mathbb{b}}}{\d N} =
    {w^\mathbb{a}}_{\mathbb{c}} {n_{\rm T}}^{\mathbb{c}\mathbb{b}}
    + {w^\mathbb{b}}_{\mathbb{c}}{n_{\rm T}}^{\mathbb{a}\mathbb{c}} + \frac{\d w^{\mathbb{a}}_{\mathbb{c}}}{\d \ln k} \Gamma^{\mathbb{c}\mathbb{b}}+ \frac{\d {w^\mathbb{b}}_{\mathbb{c}}}{\d \ln k}\Gamma^{\mathbb{a}\mathbb{c}}
+ \cdots  .
\end{equation}

\newpage
\begin{tcolorbox}
\small
\subsection{\emph{Mathematica} implementation}
In these grey panels
we discuss the numerical \emph{Mathematica}
implementation of the transport method,
available from \href{https://transportmethod.wordpress.com/}{transportmethod.com}.
In the control panel of this notebook
one can specify the model to be evaluated
and select which computations to perform.

\paranodot{Which observables?}%
By default, the notebook computes observables at a chosen scale $k$.
Computing the spectral index at this scale
using Eq.~\eqref{nab} on subhorizon scales requires
delicate cancelations between oscillating terms.
For some models it
can be considerably slower than computing $\Sigma^{ab}$ alone.
In these cases it can be preferable to evaluate $\Sigma^{ab}$ a few times
(two is typically sufficient) and compute the spectral index \emph{via}
finite difference.%
	\footnote{For practical purposes, however,
	it can be useful to solve Eq.~\eqref{nab}
	in order
	to understand how far inside the horizon we should start our
	computation; this will be discussed in {\S}\ref{sec:howdeep}.
	It is possible to select which method is used from the control panel.}
	
\para{Power spectrum}%
From the control panel one can also specify a range of $k$-scales
at which to compute $\Sigma^{ab}$,
in order to obtain the power spectrum as a function of scale $P_{\zeta}(k)$. 

\para{Example model}%
Throughout this paper we use the 3-field model
`\textsf{number 2}'
as an example.
This is a simple extension of the case studied in Ref.~\cite{1010.3693},
which
is a model of quasi-single-field inflation giving rise to a feature in
$P_{\zeta}(k)$ as a result of excitation of a heavy field via a nontrivial metric.
(See \S\ref{sec:IC} for a brief explanation of quasi-single-field scenarios.)

\textsf{Number 2} can be regarded
as a `quasi-two-field' example.
We add another light field to obtain, in addition to the nontrivial behaviour
coming from $G^{\alpha\beta}$, superhorizon evolution \emph{via} a turn
in the plane of the two light directions. This turn arrises in the region of field space where the metric is approximately the unit matrix. The turn occurs due to the hierarchy in the masses of the displaced fields --- as the heavier field approaches itÕs minimum, the direction of steepest descent becomes progressively more aligned with the lightest field.
There is no direct motivation for this model, but at a qualitative level similar
 characteristics can arise in supergravity. Here we merely employ this
 example for illustrative purposes.

The model has an equation of motion of the form~\eqref{eq:KG}.
The potential is
\begin{equation}
V=\frac{1}{2}\sum_{\alpha=1}^{3}m_{\alpha}^{2}\phi_{\alpha}^{2}
\end{equation}
and the mass ratios are
$m_{2}^2/m_{1}^{2}=30$
and $m_{3}^2/m_{1}^{2}=1/81$.
The metric takes the form
\begin{equation}
G^{\alpha\beta}=
\begin{pmatrix}
1 & \Gamma & 0 \\
\Gamma& 1& 0 \\
0& 0& 1
\end{pmatrix},
\end{equation}
with
\begin{equation}
\Gamma \equiv \frac{0.9}{\cosh\left(2\frac{\phi_{1}^{2}-7}{0.12}\right)^{2}}.
\end{equation}
\end{tcolorbox}

\section{Initial conditions}
\label{sec:IC}

In a simple model
we would typically apply~\eqref{eq:transport-equation}
only in the superhorizon regime,
where
all masses are relevant
because $k/aH$ is exponentially small.
We would estimate
an initial value of $\Sigma^{ab}$ for
modes which are `light' in the sense that
$k/aH$ dominates ${M^\alpha}_\beta / H^2$
around horizon exit,
and set all correlation functions to zero for `heavy' modes
to which this does not apply.
The justification is two-fold.
First, heavy modes are orthogonal to the inflationary trajectory,
so if this remains nearly straight throughout the epoch
of horizon exit then fluctuations in these
directions have no physical effect.
Second, quantum fluctuations in massive
modes decay exponentially, so they become irrelevant
almost immediately.

Under certain circumstances
this approximation
may miss effects from modes with intermediate masses
of order the Hubble scale or slightly larger.
If bending of the inflationary trajectory is not negligible
during horizon exit then fluctuations in massive modes
can be partially converted into the adiabatic
density perturbation
before they have time to decay.
Chen \& Wang called this scenario
`quasi-single field inflation'~\cite{hep-th/0605045, 0909.0496,0911.3380,1010.3693,1104.1323,1211.1624,1306.5680,1307.7110,1404.7522,1404.1536,1405.4257}.
It yields a distinctive bispectrum.
Even if intermediate-mass modes are not relevant,
turns in the field-space trajectory
will cross-correlate fluctuations
in different species.
This can have a significant impact on the later
evolution of observables.
A striking example is the `destructive interference'
observed by McAllister, Renaux-Petel \& Xu in Ref.~\cite{1207.0317}.

As described
in~{\S}\ref{sec:mass-capture},
we should
account for all these effects by setting initial
conditions sufficiently early that ${M^\alpha}_\beta/H^2$
is negligible compared to $(k/aH)^2$.
Because the transport equation requires no approximation
to be made regarding ${M^\alpha}_\beta$ the subsequent
evolution is exact (at tree level)
and will capture all quasi-single field and
cross-correlation effects.

In this section we explain how initial conditions can be
computed in the deeply subhorizon regime,
where all correlation functions are dominated by kinetic contributions.

\subsection{Deep inside the horizon}
\label{sec:subhorizon}

In this section we work in conformal time, defined by
$\tau = \int^t_\infty \d t' / a(t')$.

\para{Field correlation function}%
In the massless limit,
the Feynman two-point function evaluated in the vacuum state
is%
    \footnote{The equal-time two-point function in a model
    with nontrivial field-space metric was calculated
    by Sasaki \& Stewart~\cite{Sasaki:1995aw}.
    The factor of the parallel propagator
    appearing in the unequal-time propagator
    was given in Ref.~\cite{1208.6011}.}
\cite{1208.6011}
\begin{equation}
    \langle
        \delta\phi^\alpha(\vect{k}', \tau')
        \delta\phi^\beta(\vect{k}, \tau)
    \rangle
    =
    (2\pi)^3 \delta(\vect{k} + \vect{k}')
    \frac{\Pi^{\alpha\beta}(\tau',\tau)}{2k^3}
    H(\tau) H(\tau')
    (1 - \im k \tau)
    (1 + \im k \tau')
    \e{\im k (\tau - \tau')}
    ,
    \label{eq:unequal-time-field-field}
\end{equation}
where we have assumed $\tau < \tau'$.
The quantity $\Pi^{\alpha\beta}(\tau',\tau)$ is the parallel
propagator evaluated on the inflationary trajectory,
\begin{equation}
    \Pi^{\alpha\beta}(\tau', \tau)
    \equiv
    \PathOrder
    \exp
    \bigg(
        {-\int_{\tau}^{\tau'}} \d \tau'' \;
        \Gamma^\alpha_{\lambda \mu}\big[ \phi^\nu(\tau'') \big]
        \frac{\d \phi^\lambda}{\d \tau''}
    \bigg)
    G^{\mu \beta}(\tau) .
\end{equation}
It transforms as a bitensor;
the index $\alpha$ transforms as a tensor in the tangent
space at $\phi^\nu(\tau')$,
whereas the index $\beta$ transforms as a tensor in the tangent
space at $\phi^\nu(\tau)$.
The symbol $\PathOrder$ denotes path-ordering
and rewrites its argument in order of decreasing
time along the trajectory.

We take the equal-time limit $\tau = \tau'$
in which the parallel propagator reduces to the metric,
and evaluate
the correlation function well inside the horizon where
$|k / aH| \approx |k\tau| \gg 1$.
(If $\tau$ represents a time $N$ e-folds prior to horizon
exit for the mode $k$, then $|k\tau| \approx \e{N}$.)
This yields
\begin{equation}
    \langle
        \delta\phi^\alpha(\vect{k})
        \delta\phi^\beta(\vect{k}')
    \rangle_\tau
    \approx
    (2\pi)^3 \delta(\vect{k} + \vect{k}')
    \frac{H^2 G^{\alpha\beta}}{2k^3}
    |k\tau|^2
    \label{eq:field-cf-ic}
\end{equation}
in which $H$ and $G^{\alpha\beta}$ should be evaluated
at the common time $\tau$.

\para{Vacuum state}%
Eq.~\eqref{eq:field-cf-ic}
gives a suitable initial condition
for \emph{all} field--field correlation functions
at sufficiently early times,
but only if we assume that the mode of
wavenumber $\vect{k}$ was
practically in its vacuum state for at least
a few e-folds
before the time $\tau$
so that vacuum initial conditions
were applicable.

This is not guaranteed.
If inflation is sufficiently prolonged,
a mode with fixed comoving wavenumber $\vect{k}$
must have originated on very small physical
scales.
Physics at
these scales is presumably not governed by
the effective theory used to describe inflation,
so these
short-scale modes
will join it only when they
are redshifted within its purview.
Their state at that time should properly be
regarded as a boundary condition needed to define
the effective field theory.
Like all details of ultraviolet physics, it
cannot be predicted
from within the effective theory.

The influence of such boundary conditions
was studied by
Anderson, Molina-Paris \& Mottola~\cite{Anderson:2005hi}.
They found that, to be consistent with Einstein gravity
as a low-energy description, the effective
stress tensor
generated by unobserved high-energy fluctuations
should correspond to sufficiently depopulated
occupation numbers at large frequencies.
If these occupation numbers are conserved
during redshifting then modes with these frequencies
would join the effective description
while practically in their vacuum state.
In that case Eq.~\eqref{eq:field-cf-ic}
will apply.
This is the default assumption in many inflationary models.

An alternative, studied by a number of authors
(see eg. Refs.~\cite{Anderson:2005hi,Meerburg:2009ys}),
is that modes join the effective description
at a fixed time before horizon exit
with nonzero occupation number.
In this case Eq.~\eqref{eq:field-cf-ic}
would require corrections.
Whether this occurs is a model-dependent
question, but if corrections are necessary
they can be accommodated as a change of initial
conditions.
The transport equations themselves do not require
modification.


\para{Correlation functions with momenta}%
To compute correlation functions involving momenta
we use the relation $\d N=-\d\tau/\tau$,
which implies $\curlDN = - \tau \curlDtau$.
Differentiating the unequal-time
field--field correlation function~\eqref{eq:unequal-time-field-field}}
and using that
the parallel propagator is covariantly constant,
\begin{equation}
    \curlDtauprime \Pi^{\alpha\beta}(\tau', \tau) =
    \curlDtau \Pi^{\alpha\beta}(\tau', \tau) = 0 ,
\end{equation}
we obtain the unequal-time field--momentum correlation functions,
\begin{align}
    \langle
        \delta \pi^\alpha(\vect{k}', \tau')
        \delta \phi^\beta(\vect{k}, \tau)
    \rangle
    & =
    -(2\pi)^3
    \delta(\vect{k} + \vect{k}')
    \frac{\Pi^{\alpha\beta}(\tau',\tau)}{2k^3}
    H(\tau) H(\tau')
    |k\tau'|^2
    (1 - \im k \tau)
    \e{\im k (\tau - \tau')}
    \\
    \langle
        \delta \phi^\alpha(\vect{k}', \tau')
        \delta \pi^\beta(\vect{k}, \tau)
    \rangle
    & =
    -(2\pi)^3
    \delta(\vect{k} + \vect{k}')
    \frac{\Pi^{\alpha\beta}(\tau', \tau)}{2k^3}
    H(\tau) H(\tau')
    |k\tau|^2
    (1 + \im k \tau)
    \e{\im k (\tau - \tau')} .
\end{align}
We have
neglected terms suppressed by
the slow-roll parameter $\epsilon = -\dot{H}/H^2$,
which are generated by differentiation of $H$.
In~\eqref{eq:unequal-time-field-field} we did not
retain slow-roll suppressed contributions
(although this can be done),
so retaining them here would give an inconsistent
set of corrections.
Also, we have again assumed $\tau < \tau'$.
The case $\tau > \tau'$ can be obtained from these
expressions by complex-conjugation.

The equal-time limit can be obtained as before,
but unlike the field--field correlation function
the result is complex.
However, the imaginary part vanishes at late times
and therefore does not affect observables.
For simplicity we can work with the symmetrized
(or `Weyl ordered') correlation function
which is always real and coincides with the other
field--momentum two-point functions outside the horizon,
\begin{equation}
    \frac{1}{2}
    \big\langle
        \delta \pi^\alpha(\vect{k})
        \delta \phi^\beta(\vect{k}')
        +
        \delta \phi^\beta(\vect{k}')
        \delta \pi^\alpha(\vect{k})
    \big\rangle_\tau
    =
    - (2\pi)^3
    \delta(\vect{k} + \vect{k}')
    \frac{H^2 G^{\alpha\beta}}{2k^3}
    |k\tau|^2 .
\end{equation}
By a very similar procedure, the
momentum--momentum correlation function
can be found (also to leading order
in slow-roll terms) to be
\begin{equation}
    \langle
        \delta \pi^\alpha(\vect{k})
        \delta \pi^\beta(\vect{k}')
    \rangle
    =
    (2\pi)^3
    \delta(\vect{k} + \vect{k}')
    \frac{H^2 G^{\alpha\beta}}{2k^3}
    |k\tau|^4 .
\end{equation}
Collecting these results yields the
\emph{universal} initial conditions
\begin{equation}
\begin{aligned}
    \Sigma^{\alpha\beta}_\ast
    & =
    \frac{H^2_\ast G^{\alpha\beta}_\ast}{2}   |k\tau_\ast|^2  ,
    &
    \Sigma^{\bar{\alpha}\beta}_\ast
    =
    \Sigma^{\alpha\bar{\beta}}_\ast
    & =
    - \frac{H^2_\ast G^{\alpha\beta}_\ast}{2}   |k\tau_\ast|^2 ,
    &
    \Sigma^{\bar{\alpha}\bar{\beta}}_\ast
    & =
    \frac{H^2_\ast G^{\alpha\beta}_\ast}{2}   |k\tau_\ast|^4  ,
    \\
    n_\ast^{\alpha\beta}
    & =
    H^2_\ast G^{\alpha\beta}_\ast \  |k\tau_\ast|^2 ,
    &
    n^{\bar{\alpha}\beta}_\ast
    =
    n^{\alpha\bar{\beta}}_\ast
    & =
    - H^2_\ast G^{\alpha\beta}_\ast   |k\tau_\ast|^2  ,
    &
    n^{\bar{\alpha}\bar{\beta}}_\ast
    & =
    2 H^2_\ast G^{\alpha\beta}_\ast   |k\tau_\ast|^4 
\end{aligned}
\label{eq:universal-ics}
\end{equation}
where a subscript `$\ast$' denotes evaluation at the initial time.

As tensor modes behave like free scalar fields
(apart from a change in normalization)
their initial conditions follow immediately,
\begin{equation}
\begin{aligned}
    \Gamma_{\ast}^{\gamma\gamma}
    & =
    \frac{H^2_\ast}{\Mp^2}   |k\tau_\ast|^2  ,
    &
    \Gamma_{\ast}^{\pi\gamma}
    =
    \Gamma_{\ast}^{\gamma\pi}
    & =
    - \frac{H^2_\ast }{\Mp^2}   |k\tau_\ast|^2 ,
    &
    \Gamma^{\pi\pi}_\ast
    & =
    \frac{H^2_\ast }{\Mp^2}   |k\tau_\ast|^4
    \\
  n_{{\rm T}\ast}^{\gamma\gamma}
    & =
    \frac{2 H^2_\ast}{\Mp^2}   |k\tau_\ast|^2  ,
    &
    n_{{\rm T}\ast}^{\pi\gamma}
    =
    n_{{\rm T}\ast}^{\gamma\pi}
    & =
    - \frac{2 H^2_\ast }{\Mp^2}   |k\tau_\ast|^2 ,
    &
    n_{{\rm T}\ast}^{\pi\pi}
    & =
    \frac{4 H^2_\ast }{\Mp^2}   |k\tau_\ast|^4.
\end{aligned}
\end{equation}

\para{Slow-roll corrections}%
In principle, the initial conditions~\eqref{eq:universal-ics}
should be corrected by slow-roll
terms proportional to powers of $\epsilon$ or its derivatives.
Therefore although the transport equation~\eqref{eq:transport-equation}
makes no use of the slow-roll approximation,
our use of~\eqref{eq:universal-ics}
\emph{does} require that slow-roll is a fair approximation
near the initial time.
We do not need any form of the slow-roll approximation thereafter;
the slow-roll conditions may be badly violated
or fail entirely.

In certain cases it may happen that a solution with initial conditions
chosen to satisfy Eq.~\eqref{eq:universal-ics}
will still relax to the correct solution, even if slow-roll
is only marginally valid
or weakly violated
near the initial time,
provided we begin the calculation sufficiently far before horizon exit.
As for any numerical solution, some care may be required
to check that results are stable to changes in the grid and
the initial time.
In practical calculations this implies that the initial time should
be chosen so that it is comfortably earlier
than any interesting dynamical effects which we
hope to capture.
\\

\begin{tcolorbox}
\small
\subsection{\emph{Mathematica} implementation}
\label{sec:howdeep}
Increasing the number of e-folds of subhorizon evolution slows down the solver;
but not starting the calculation sufficiently early leads to inaccurate results. 
The amount of subhorizon evolution (as well as the accuracy settings of the Mathematica function NDSolve) required to obtain a sufficiently accurate result is model dependent, however it is useful to have a prescription that works as a guideline.

One way to approach this is to use the behaviour of the spectral index.
Sufficiently far inside the horizon, the dominant contribution to the scale dependence
comes from $\Sigma^{\bar{\alpha}\bar{\beta}}_0 \propto k^4$.
Therefore we expect the spectral index $n_s -1$ to be roughly 4 at early times,
independent of the model. 
If the system is not given enough e-folds of subhorizon evolution,
the delicate cancelations that occur for $n_{ab}$ will quickly break and
give $n_s -1\ne 4$.
We have found that ensuring $n_s -1= 4$ for a sustained period of order $\sim 3$
e-folds is usually sufficient to obtain consistently good results. This method should not be used as a replacement for testing for convergence against changes in initial time and grid, but it can be used as an indicator.
\end{tcolorbox}

\section{Connection with observables}

\subsection{From field space to $\zeta$}
\label{sec:gauge}

To calculate observables,
the flat-gauge field and momentum correlation
functions
must be converted
to correlation functions of the
uniform-density gauge
curvature perturbation $\zeta$.
Assuming all isocurvature modes decay,
it is the curvature perturbation which sets
initial conditions for density fluctuations
in the later universe.
In this section we briefly explain
how an appropriate transformation
can be extracted from the separate universe assumption
using the methods of Ref.~\cite{1410.3491}.
As explained there, the gauge transformation could
equally well be derived from traditional
perturbation theory in the large-scale limit.

\para{Gauge transformation}%
According to the construction of Ref.~\cite{1410.3491},
the number of e-folds $\Delta N$ between a point $p$ on a spatially
flat hypersurface at which the density is $\rho_p$
and an equivalent point on a nearby uniform density
hypersurface with density $\rho_\ast$ is
\begin{equation}
    \Delta N =
    \left.
    \frac{\d N}{\d \rho}
    \right|_p
    (\rho_\ast - \rho_p)
    + \cdots ,
\end{equation}
where the omitted terms are higher order in $\rho_\ast - \rho_p$.
Under a variation of $p$, it follows that
\begin{equation}
\begin{split}
    \delta ( \Delta N )
    & \approx
    {-\mbox{}}
    \left.
    \frac{\d N}{\d \rho}
    \right|_p
    \delta \rho_p
    + \cdots
    \approx
    {-\mbox{}}
    \left.
    \frac{\d N}{\d \rho}
    \right|_p
    \Big(
        \left. \frac{\partial \rho}{\partial \phi^\alpha} \right|_p \delta \phi^\alpha_p
        +
        \left. \frac{\partial \rho}{\partial \pi^\alpha} \right|_p \delta \pi^\alpha_p
    \Big)
    + \cdots
    \\ &
    \equiv
    N_\alpha \delta \phi^\alpha
    +
    N_{\bar{\alpha}} \delta \pi^\alpha
    + \cdots ,
\end{split}
\label{eq:gauge-xfm}
\end{equation}
where `$\cdots$' denotes terms of higher order in
$\delta \phi^\alpha$ or $\delta \pi^\alpha$
which are not needed for the first-order
gauge transformation.
The last equality should be interpreted as a definition
of $N_\alpha$ and $N_{\bar{\alpha}}$.
No use is being made of the slow-roll approximation
so the density $\rho$ contains both potential and
kinetic contributions.
Performing the partial derivatives, we find
\begin{subequations}
\begin{align}
    \label{eq:N-unbarred}
    N_\alpha
    & =
    \frac{1}{2\epsilon}
    \frac{V_\alpha}{V} \\
    \label{eq:N-barred}
    N_{\bar{\alpha}}
    & =
    \frac{1}{2\epsilon(3-\epsilon)} \frac{\pi_\alpha}{\Mp^2} .
\end{align}
\end{subequations}
The variation $\delta ( \Delta N )$ gives the fluctuation in
e-folds required to reach $\rho_\ast$,
and therefore must be the curvature perturbation $\zeta$.
It follows that
Eq.~\eqref{eq:gauge-xfm}
expresses the gauge transformation to $\zeta$ at linear order.

This argument was given in Ref.~\cite{1208.6011} to lowest
order in the slow-roll approximation
where the contribution for $\delta \pi^\alpha$ can be neglected.
In the formulation given here, Eqs.~\eqref{eq:gauge-xfm}
and \eqref{eq:N-unbarred}--\eqref{eq:N-unbarred}
apply to all orders in the slow-roll expansion.
The argument of Ref.~\cite{1410.3491},
generalized to a nontrivial field-space metric,
shows that if desired we can use the Hamiltonian constraint
to eliminate the $\delta \pi^\alpha$ term. This would give
\begin{subequations}
\begin{align}
    \label{eq:N-unbarred-simple}
    N_\alpha
    & = - \frac{1}{2 \epsilon} \frac{\pi_\alpha}{\Mp^2} \\
    \label{eq:N-barred-simple}
    N_{\bar{\alpha}}
    & = 0 .
\end{align}
\end{subequations}
Like~\eqref{eq:N-unbarred}--\eqref{eq:N-barred},
Eqs.~\eqref{eq:N-unbarred-simple}--\eqref{eq:N-barred-simple}
do not invoke the slow-roll approximation.

\para{Power spectrum}%
The quantity of principal interest is the power spectrum, which is defined
in terms of the equal time $\zeta\zeta$ two-point function,
\begin{equation}
    \langle
        \zeta(\vect{k}_1)
        \zeta(\vect{k}_2)
    \rangle
    =
    (2\pi)^3
    \delta(\vect{k}_1+\vect{k}_2)
    \frac{P_{\zeta}}{k^3} .
\end{equation}
In terms of the
flat-gauge correlation functions, it follows from
Eq.~\eqref{eq:gauge-xfm} that
$P_\zeta$ can be written
\begin{equation}\label{eq:2p}
    P_{\zeta}=N_{a}N_{b}\Sigma^{a b}.
\end{equation}
The $k$-dependence of the power spectrum can be handled similarly.
We define the scalar spectral index to satisfy
\begin{equation}
    n_{s}-1 \equiv \left. \frac{\d \ln{P_{\zeta}}}{\d \ln{k}} \right|_{k = k_\star},
\end{equation}
where $k_\star$ is the pivot scale.
(The subscript `$\star$' representing evaluation at the pivot scale
should not be confused with the subscript `$\ast$' denoting
evaluation at the initial time in Eqs.~\eqref{eq:universal-ics}.)
We find
\begin{equation}
	n_{s}-1 =
		\frac{N_{a}N_{b}}{P_{\zeta}}
		\frac{\d \Sigma^{ab}}{\d \ln{k}}
		= \frac{N_{a}N_{b} n^{ab}}
			{N_{c} N_{d} \Sigma^{cd}} .
	\label{eq:ns}
\end{equation}
The running of the spectral index is defined to be
\begin{equation}
    \alpha \equiv \left. \frac{\d n_s}{\d \ln{k}} \right|_{k = k_\star} .
\end{equation}
It
can be computed using a
finite-difference approximation, although another
transport equation could be written for it if desired.

\para{Tensor fraction}%
The tensor power spectrum is defined by analogy with the scalar power spectrum
\begin{equation}
    \langle
        \gamma_{ij}(\vect{k}_1)
        \gamma_{ij}(\vect{k}_2)
    \rangle
\equiv
    (2\pi)^3
    \delta(\vect{k}_1+\vect{k}_2)
    \frac{P_\gamma}{k^3} .
\end{equation}
The power in each polarization adds incoherently.
Using the normalization condition
$e^s_{ij}e^{s'}_{ij}=2\delta^{ss'}$
it follows that the total tensor power satisfies
\begin{equation}
    \langle
        \gamma_{ij}(\vect{k}_1)
        \gamma_{ij}(\vect{k}_2)
    \rangle = \sum_s \sum_{s'}  \langle
        \gamma_s(\vect{k}_1)
        \gamma_s(\vect{k}_2)
    \rangle e^s_{ij}e^{s'}_{ij}=
    2 \sum_s \langle
        \gamma_{s}(\vect{k}_1)
        \gamma_{s}(\vect{k}_2) \rangle .
\end{equation}
Each polarization is the same and therefore the final
result is $4 \langle \gamma_+ \gamma_+ \rangle$,
or equivalently $4 \langle \gamma_\times \gamma_\times \rangle$.
The tensor-to-scalar ratio $r$ is
defined to be
\begin{equation}
  r \equiv \frac{P_\gamma}{P_\zeta} = \frac{4 \Gamma^{\gamma\gamma}}{P_{\zeta}}
\end{equation}
and the tensor spectral index is
\begin{equation}
    n_{\rm T}= \left. \frac{\d \ln{P_{\gamma}}}{\d \ln{k}} \right|_{k = k_\star} = \frac{n_T^{\gamma\gamma}}{\Gamma^{\gamma\gamma}}.
\end{equation}

\subsection{End of inflation or beyond?}
\label{sec:adiabatic}

To complete the calculation, we must decide
when to terminate the integration
and measure final values for each observable.

In principle we should track the evolution of all correlation functions up
to the last scattering surface---%
or until just before horizon re-entry if we wish to
study the assembly of large-scale structure.
In practice this is extremely challenging,
partly because the results depend on the unknown details of reheating
and partly because we have little direct knowledge of the epochs between reheating
and horizon re-entry.
(For recent literature studying the impact of reheating on
inflationary observables, see Refs.~\cite{1206.5196, 1303.4678, 1311.3972,1405.6195}.
Ref.~\cite{1410.3808} is a recent review of reheating in general.)

It is possible for $\zeta$ to evolve whenever power
remains in any `isocurvature' modes, by which we mean phase space directions
transverse to the inflationary trajectory.
Microwave background data now strongly constrain the presence of isocurvature
modes around the time of photon decoupling at $z \sim 1100$,
but this provides only a lower limit on the decay time.
We would normally aim to terminate the integration as early as is safe,
to avoid being obliged to integrate through periods of the universe's history
where we must make assumptions about its evolution.
This means we must be able to determine when all power in isocurvature
modes has become exhausted---what is called the `adiabatic limit'~\cite{GarciaBellido:1995qq,Elliston:2011dr}.

\para{Isocurvature power during inflation}%
First consider evolution during the inflationary era.
One option would be to track the two-point functions of each relevant
degree of freedom, using Gram--Schmidt orthogonalization to
construct linear combinations which measure the power transverse
to the inflationary trajectory.
For practical purposes we could suppose that the system is close
enough to an adiabatic limit
whenever all of these two-point functions become sufficiently small.

This approach is feasible, but rather cumbersome.
At least during slow-roll evolution an alternative is
to use the optical analogy of trajectories flowing over field-space
developed in Ref.~\cite{Seery:2012vj}.
When slow-roll is a good approximation there is no need to
track momentum perturbations, and we can write 
an equation analogous to~\eqref{eq:expansion-tensor}
purely for the field fluctuations,
\begin{equation}
	\curlDN \delta\phi^\alpha = {w^\alpha}_{\beta} \delta \phi^\beta ,	
\end{equation}
where ${w^\alpha}_{\beta}$ is likewise an analogue
of the expansion tensor~\eqref{eq:u-tensor},
\begin{equation}
	{w^\alpha}_{\beta}
	=
		\curlDplain_\beta(\curlDN \phi^\alpha)
		- \frac{1}{3} {R^\alpha}_{\gamma\lambda\beta} \curlDN \phi^\gamma \curlDN \phi^\lambda
	=
		\curlDplain_\beta \left(-\frac{V^{\alpha}}{3H^2}\right)
		- {R^\alpha}_{\gamma\lambda\beta} \frac{V^{\gamma}V^{\lambda}}{27 H^4} .
\end{equation}
Here, $V^\alpha = G^{\alpha\beta} V_\beta$.
As explained in Ref.~\cite{Seery:2012vj},
the eigenvalues of ${w^\alpha}_\beta$
can be used to detect the presence of growing
and decaying modes:
a positive eigenvalue indicates a growing mode,
whereas a negative eigenvalue indicates a decaying one.

In most circumstances one mode is constant or slowly evolving,
and therefore gives an eigenvalue which is zero or
slightly positive.
Therefore, in an $N$-field system,
approach to the adiabatic limit is signalled
by the appearance of $N-1$ large negative eigenvalues.
This method is simpler than computing all combinations
of isocurvature correlation functions, but
clearly shares its arbitrariness in
deciding when the fluctuations have decayed sufficiently
to declare that an adiabatic limit has been reached.
There is always the possibility that extremely violent future
dynamics could amplify even very small modes.

\begin{figure}
    \centering \includegraphics[width=0.8\textwidth]{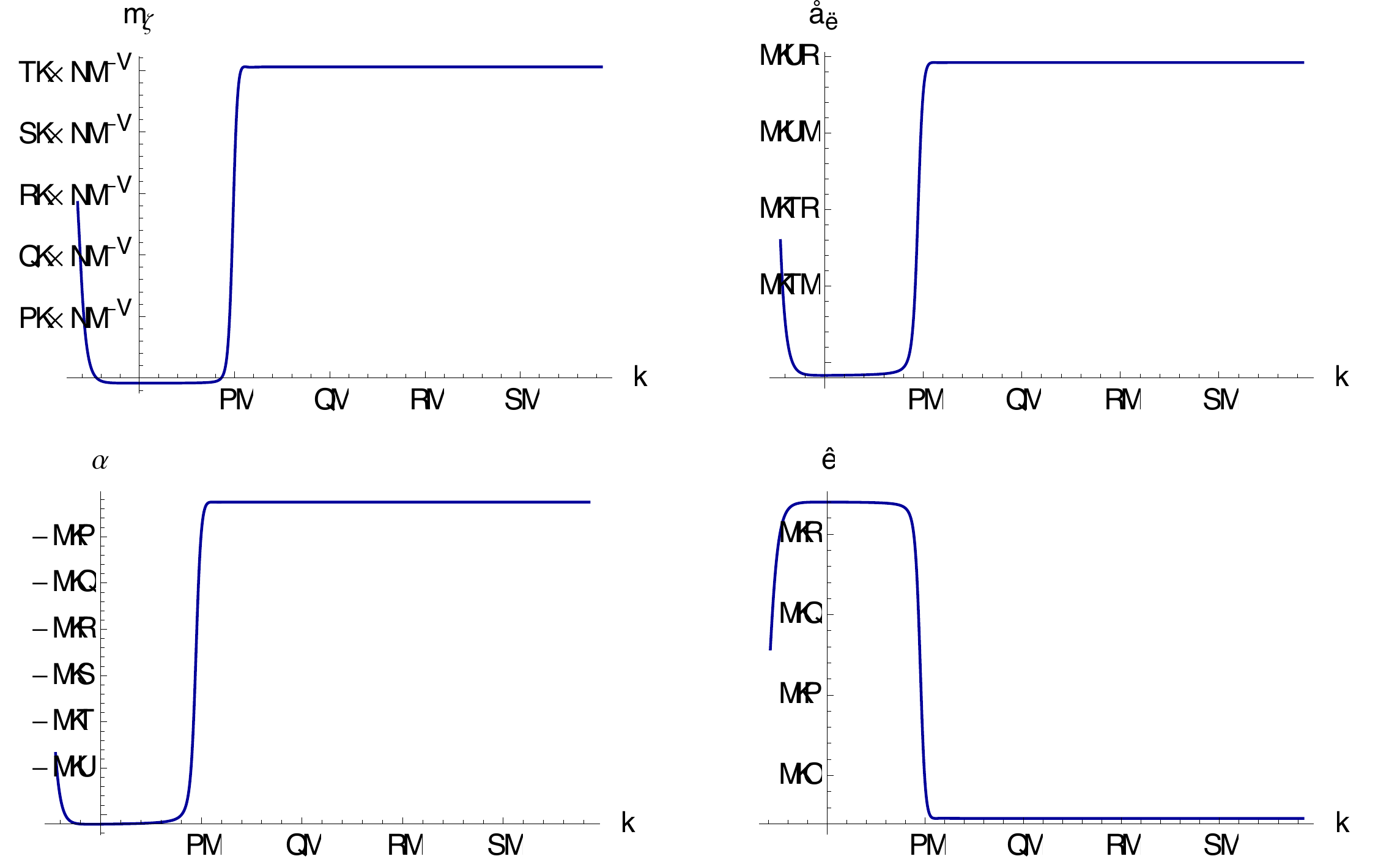} 
    \caption{Evolution of observables as a function of e-folds $N$:
    clockwise from top left,
    the power spectrum $P_{\zeta}$;
    the spectral index $n_s$;
    the running of spectral index $\alpha$;
    and the tensor-to-scalar ratio $r$.}
    \label{fig:observablesofN}
\end{figure}

\para{After inflation}%
This test applies only during slow-roll evolution, and therefore
will typically become unreliable some time \emph{before} the
end of inflation unless this is mediated by a sudden event such as
a waterfall transition.
When it applies, however, it may provide a rationale for
terminating the integration at or before the end of the slow-roll phase.
This is the best possible outcome.

Much less can be said if slow-roll breaks down before
complete decay of the isocurvature modes.
In this case, one should follow the decay of the scalar
species relevant during inflation into reheating products.
Isocurvature modes may be transferred or amplified during this
process.
One must then begin a second integration,
following the evolution of these fluctuations using
suitable phase space coordinates; normally,
the scalar species supporting inflation will no
longer be the relevant variables,
and the transport equation
for their correlation functions will
need to be replaced.
The range of phenomenology
which can occur during this post-inflationary phase is comparatively
under-explored, and almost certainly model-dependent.
\newpage

\begin{figure}
    \centering \includegraphics[width=0.8\textwidth]{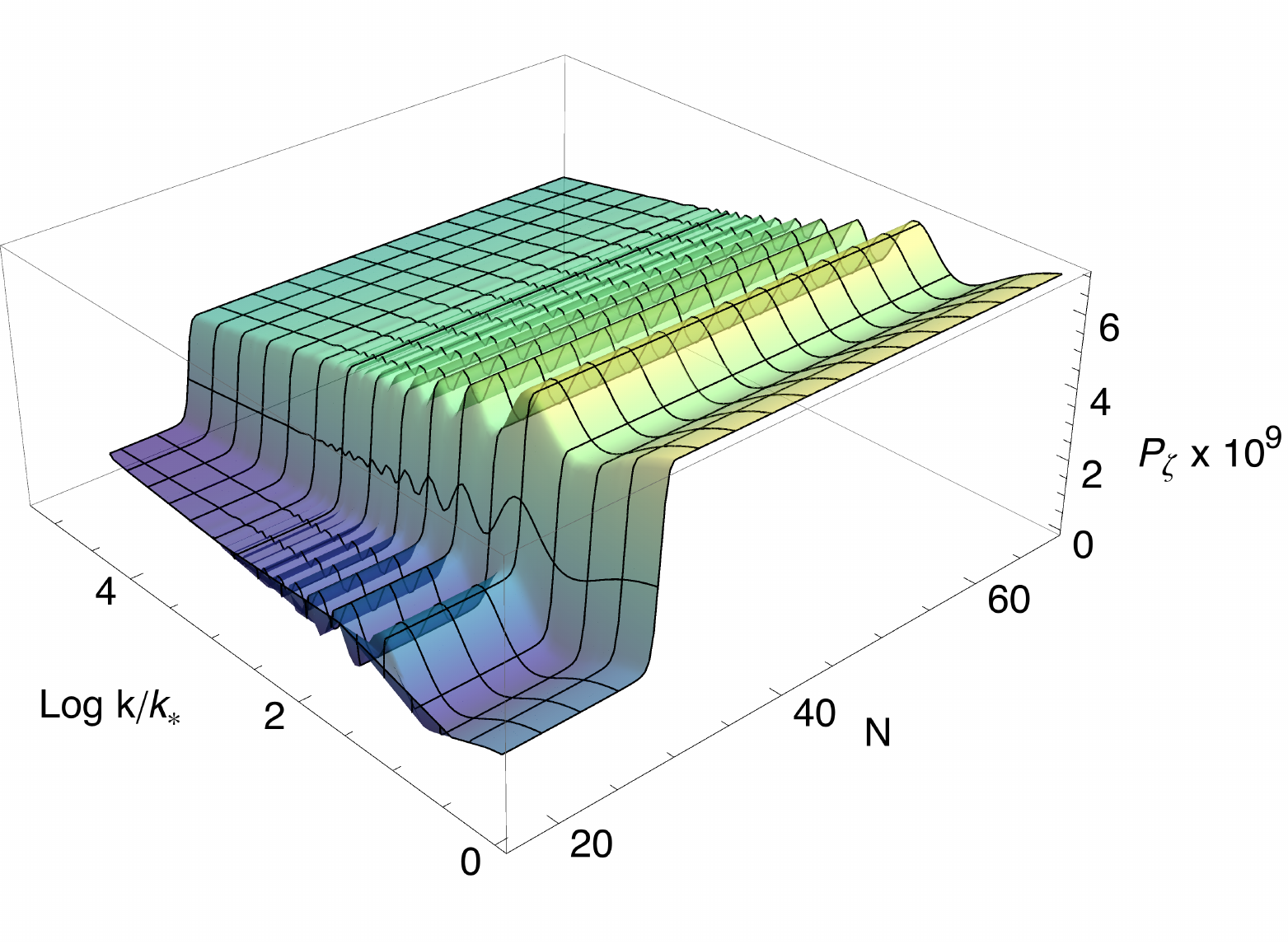} 
    \caption{The power spectrum $P_{\zeta}$ as a function of e-folds $N$ and wavenumber $k$.
    The range of scales corresponds to about 5 e-folds.
    Note the step in $P_{\zeta}(N)$ corresponding to superhorizon evolution,
    and the oscillatory features in $P_{\zeta}(k)$ as a result of
    excitations of the heavy mode around horizon exit.}
    \label{fig:PofNk}
\end{figure}

\begin{figure}[t]
\centering \includegraphics[width=0.405\textwidth]{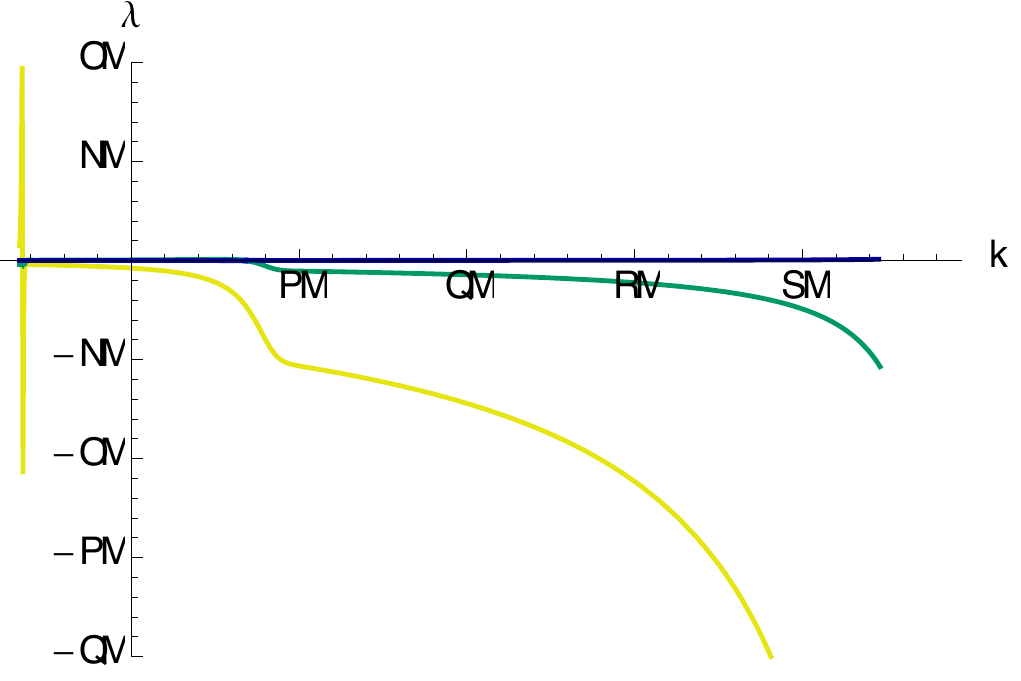} 
\centering \includegraphics[width=0.405\textwidth]{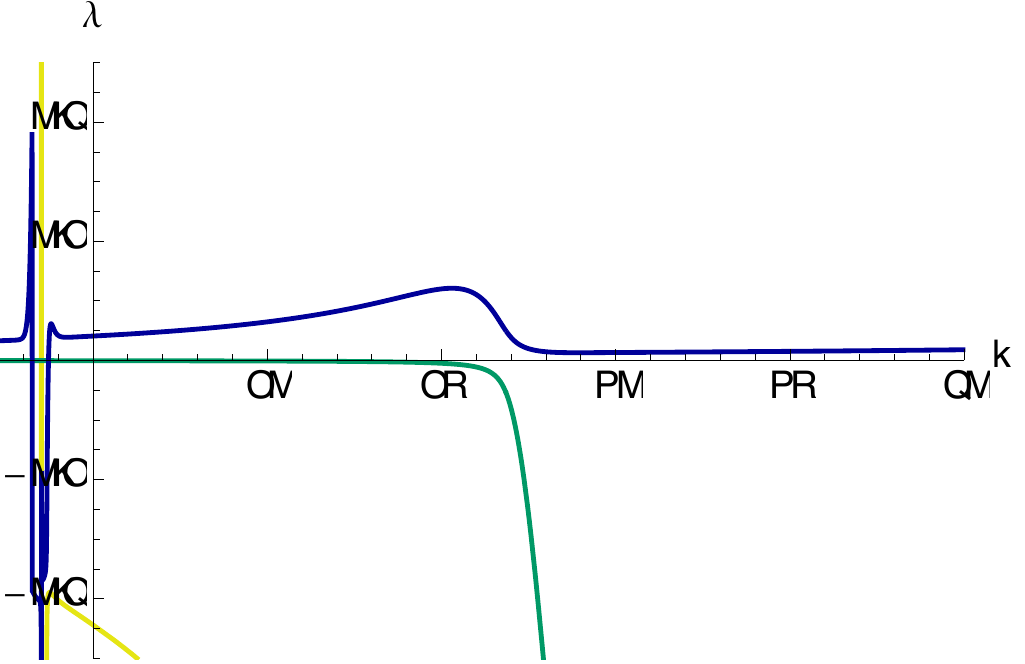} 
 \caption{Eigenvalues of the slow-roll ${u^\alpha}_{\beta}$ as a function of number of e-folds. The left plot shows the whole evolution from horizon exit to the end of inflation. The right plot is zoomed in order to understand the transfer of power around the turn in field space.}
 \label{fig:evaluesofN}
\end{figure}

\begin{tcolorbox}
\small
\subsection{\emph{Mathematica} implementation}

As above, we use the model \textsf{Number 2} as an illustration.

\para{Time dependence}%
In
Fig.~\ref{fig:observablesofN} we
plot the
time evolution of the power spectrum, spectral index, running of the spectral
index and tensor-to-scalar ratio for the scale leaving the horizon
55 e-folds before the end of inflation.

A turn in field space occurs after $N\sim30$ e-folds.
The turn causes isocurvature modes to source evolution of $P_{\zeta}$,
visible here as a large step.
This corresponds to a sudden jump in the spectral index and running,
and a corresponding drop in the value of the tensor-to-scalar ratio.
Once the trajectory settles into the valley of the potential,
no further evolution occurs. 

\para{Scale dependence}%
In
Fig.~\ref{fig:PofNk} we plot the time and wavenumber
evolution of the function $P_{\zeta}(k)$.
Evidently each scale undergoes qualitatively similar evolution to the
example shown in Fig.~\ref{fig:observablesofN}.
There is a significant step around $N\sim30$.
However, it is also possible to see the effect of excitation
of the heavy mode: it induces oscillatory features in $P_{\zeta}(k)$ as
a function of wavenumber.

\para{Approach to the adiabatic limit}%
We plot the time evolution of the eigenvalues
of the slow-roll expansion tensor ${w^\alpha}_\beta$ in
Fig.~\ref{fig:evaluesofN}.
Around horizon crossing the metric is designed to excite the heaviest field, leading to a 
sudden enhancement of the isocurvature modes,
here visible as a sharp excursion to positive eigenvalues.

Soon after, one of the eigenvalues (the yellow line) becomes much more negative
than the other two, indicating that isocurvature fluctuations associated with
the heaviest field are decaying exponentially.
The suppression is so rapid that the system is subsequently well-approximated
by a two-field model.

After 20 e-folds of superhorizon evolution the trajectory turns,
causing excitation of the remaining isocurvature mode.
Its power is subsequently transferred to the adiabatic direction
(approximately represented by the blue line),
generating the step-like features in Fig.~\ref{fig:observablesofN}.
After the turn the remaining isocurvature mode
(the green line)
is rapidly suppressed.
At this point an adiabatic limit has been reached:
the system has become effectively single-field, and
$\zeta$ is conserved.
\end{tcolorbox}

\section{Final summary}

Computing precise predictions for observables in complex inflationary
models requires tools going beyond
textbook methods.
This is especially true for models with a nontrivial field-space metric.

First, it is helpful (but not mandatory)
to use a covariant description of the system,
especially when computing correlation functions of higher-order
(see, eg., Ref.~\cite{1208.6011}).
Second, to track the influence of curvature scales
associated with the metric we require a
computational method which retains information about
\emph{all} mass scales in the problem.
In this paper we describe a simple method for doing so,
beginning with universal massless initial
conditions
long before horizon exit
and solving a transport equation for the subsequent evolution.
We have focused on the equal-time two-point functions
because only these are required for the simplest inflationary
observables.
Nevertheless,
only straightforward modifications are needed
to compute
unequal-time correlation functions
or those of higher order.

This method is applicable to a large class of models,
including models descending from ideas in string theory
or supergravity where nontrivial field-space
metrics are often associated with a nontrivial K\"{a}hler
potential.
It gives a simple description which allows the analysis
to proceed from deeply subhorizon scales to
the end of inflation (or beyond), with no matching
required around the time of horizon exit.
The evolutionary equation does not
make use of the slow-roll approximation---%
although our analytic initial conditions do---%
and therefore accounts for effects from
violation of the slow-roll conditions,
turns in field-space at any point on the inflationary
trajectory,
and the influence of massive fields or quasi-single-field dynamics.
It does not yet apply to models with entirely arbitrary
kinetic terms, such as $P(X)$ or $P(X,\phi)$ models.
We leave these for future work.

In this paper we have tried to explain how the transport
method can be applied in practice to obtain
predictions from inflationary models which exhibit
one or more of these complexities.
In addition we have attempted to highlight
those points where our implementation goes beyond
standard textbook methods
such as the separate-universe approximation,
and also those situations to which the method cannot
yet be applied. In these cases we have indicated
whether the obstruction is a matter of principle,
or just an artefact of current technology.
Since our focus is on practical usage, we
have provided a complete \emph{Mathematica} implementation
which is used as an example.
The code is intended to be accessible.
We hope it will serve both as a useful tool
for investigating realistic models, and
a platform to extend the range of scenarios
for which predictions can be obtained.

\begin{acknowledgments}

We thank Kepa Sousa, Y{\hspace{-0.1em}}vette Welling and Oliver Janssen for feedback on early versions of the draft and code. 
JF is supported by IKERBASQUE, the Basque Foundation
for Science. DS acknowledges support from the
Science and Technology Facilities Council [grant number ST/L000652/1]
and the Leverhulme Trust.
The research leading to these results has received funding
from the European Research Council under the
European Union's Seventh Framework Programme
(FP/2007–2013) / ERC Grant Agreement No. [308082].
\end{acknowledgments}


\bibliography{2pfarxiv}

\end{document}